\documentclass[12pt]{article}
\usepackage[utf8]{inputenc}
\usepackage{fullpage}
\usepackage{graphicx} % support the \includegraphics command and options

\usepackage{booktabs} % for much better looking tables
\usepackage{paralist} % very flexible & customisable lists (eg. enumerate/itemize, etc.)
\usepackage{verbatim} % adds environment for commenting out blocks of text & for better verbatim
\usepackage{amsmath}
\usepackage{amsfonts}
\usepackage{amsthm}
\usepackage{array}
\usepackage{graphicx}
\usepackage{caption}
\usepackage{subcaption}
\PassOptionsToPackage{hyphens}{url}\usepackage{hyperref}
\usepackage{bm}

\theoremstyle{plain}

\newcommand{\indep}{\perp \!\!\! \perp}

\usepackage[numbers]{natbib}
\usepackage{authblk}

\title{The Impact of Time Series Length and Discretization on Longitudinal Causal Estimation Methods}
\author[1]{Roy Adams}
\author[1,2,3]{Suchi Saria}
\author[4]{Michael Rosenblum}
\affil[1]{Department of Computer Science, Johns Hopkins University, Baltimore, MD 21218, USA}
\affil[2]{Department of Applied Mathematics and Statistics, Johns Hopkins University, Baltimore, MD 21218, USA}
\affil[3]{Bayesian Health, New York, NY 10005, USA}
\affil[4]{Department of Biostatistics, Johns Hopkins Bloomberg School of Public Health, Baltimore, MD 21205, USA}

\date{}

\newif\ifinlinefigs
\inlinefigstrue

\begin{document}

\maketitle

\begin{abstract}
    The use of observational time series data to assess the impact of multi-time point interventions is becoming increasingly common as more health and activity data are collected and digitized via wearables, social media, and electronic health records. Such time series may involve hundreds or thousands of irregularly sampled observations. One common analysis approach is to simplify such time series by first discretizing them into sequences before applying a discrete-time estimation method that adjusts for time-dependent confounding. In certain settings, this discretization results in sequences with many time points; however, the empirical properties of longitudinal causal estimators have not been systematically compared on long sequences. We compare three representative longitudinal causal estimation methods on simulated and real clinical data. Our simulations and analyses  assume a Markov structure and that longitudinal treatments/exposures are binary-valued and have at most a single jump point. We identify sources of bias that arise from temporally discretizing the data and provide practical guidance for discretizing data and choosing between methods when working with long sequences. Additionally, we compare these estimators on real electronic health record data, evaluating the impact of early treatment for patients with a life-threatening complication of infection called sepsis. 

\end{abstract}

% \section{TODOs}
% \input{sections/todos}

\section{Introduction}
% The use of observational time series data to assess the impact of
% multi-timepoint interventions is becoming increasingly common as more health
% and activity data are collected and digitized via wearables, social media, and
% health records. Such time series may involve hundreds or thousands of
% irregularly sampled observations and it is typical to simplify such time series
% by first discretizing them into sequences before applying one of several
% discrete-time estimation methods that adjust for confounding. In many settings,
% this discretization results in sequences with tens or hundreds of time points.

Sequence length can have major practical implications for the performance of
discrete-time longitudinal statistical methods. For example, a well-known drawback of inverse
probability weighting (IPW)  methods \citep{horvitz1952generalization,robins2000marginal}, and of importance sampling methods more
broadly, is that estimator variance can grow exponentially with sequence length
\citep{glynn1994importance}. However, the   performance of other popular
longitudinal causal estimators has not been systematically compared on
sequences comprising tens or hundreds of time points, such as those occurring in observational health settings, e.g., \citep{neugebauer2014targeted,lim2018forecasting,komorowski2018artificial,sofrygin2019targeted}. Further, the primary means for
controlling sequence length, increasing the width of the discrete time bins, is
known to introduce bias in certain settings \citep{schulam2018discretizing}.
Thus, practitioners must balance bias and variance concerns when making
discretization decisions. We investigate this bias/variance
trade-off.

We evaluate the impact of discretization and sequence length on three popular
longitudinal causal estimators: the aforementioned IPW estimators, iterative
regression estimators based directly on the g-computation formula (IR)
(sometimes referred to as the parametric g-computation formula \citep{robins1986new}), and targeted minimum loss-based estimators (TMLE) \citep{van2006tmle,van2012targeted}. Our goal is
to provide practical guidance for choosing between estimators and making
discretization decisions. We address the question: How does
discretization bin width --- and, by extension, sequence length --- affect bias
and variance for each of these estimators? To answer this question, we first use
simulations to compare the performance of these estimators  as we
vary the discretization bin width (Section \ref{sec:synthetic_experiments}).
Next, under a Markov assumption, we analytically identify one source of potential bias incurred by the IR and TMLE methods as
a result of discretization and provide recommendations for avoiding this bias (Section \ref{sec:discretization_bias_ir}). Finally, we compare these estimators on a real clinical causal estimation problem using electronic health record (EHR) data with up to 73 discrete time points (Section \ref{sec:real_data_experiments}). 

This work builds on a growing literature examining the behavior of longitudinal
causal estimation methods.
\citet{gottesman2018evaluating} and \citet{tran2019double} both presented
comparisons of longitudinal estimation methods on simulated and
observational health data and both considered the effect of sequence length on
estimator performance; however, both studies focused on how increasing the time
horizon, and thus changing the estimand, increases the likelihood of
finite sample positivity violations as fewer participants follow the treatment regime of interest. While we comment briefly on finite sample positivity violations, our focus is on the effect of discretization decisions \emph{for a fixed estimand and without changing the number of participants who follow the treatment regime of interest}. Further, \citet{tran2019double} used fixed length sequences with up to 6 time points and \citet{gottesman2018evaluating} used variable length sequences with an average of 13 time points. In this work, we consider simulated sequences with up to 257 time points and real clinical data with up to 73 time points. \citet{sofrygin2019targeted} examined the effect of discretization decisions on a particular TMLE estimator in an observational health setting. Their analysis included sequences with up to 143 time points, but did not compare performance across estimators. The combination of these related works and our paper may provide guidance for the design and analysis of studies involving non-experimental longitudinal data.

Another area in which causal
inference methods have been applied to long sequential data is in off-policy evaluation of reinforcement
learning (RL) algorithms. Off-policy evaluation refers to estimating the
expected reward (i.e., outcome) under a particular policy of interest from
observational data that was generated according to a different policy. This is equivalent to
longitudinal causal inference from observational data. While off-policy
evaluation is frequently done using IPW-based estimators, recent work has used doubly-robust estimators in order to reduce variance. For example, \citet{jiang2016doubly} and \citet{thomas2016data} presented such doubly-robust
estimators, observing that they reduced variance relative to IPW estimators.
% \citet{thomas2016data} proposed a combination of model-based and IPW-based
% estimators where the IPW-based estimator is used to estimate reward up to some
% time $t$ and the model-based estimator is used for all rewards after time $t$
% where the combination is tuned to minimize estimator error. 
The estimators used in \citet{jiang2016doubly} and \citet{thomas2016data}
differ from those considered in this paper in that they assume certain nuisance parameters were estimated while the evaluation policy was being learned. 
\cite{RobinsOrellanaRotnitzkysim.3301} consider the  related problem of extrapolating an optimal treatment policy  from one population to a biologically similar population  where observations are made less frequently. 
% Further, it is frequently
% assumed that the observational policy is known, and therefore
% the consistency of IPW and doubly-robust methods is guaranteed.

A final important area of related work is in the development of estimators that
operate on irregularly sampled continuous time data. For example,
\citet{soleimani2017treatment} and \citet{schulam2017reliable} modeled both the treatment and feature values, as well as the timing of observations using
Bayesian non-parametric methods. An advantage of these methods is
that they obviate the need to discretize longitudinal data; however, they
require modeling the joint distribution of all time-varying covariates which
generally requires stronger modeling assumptions than here. While the focus of our work
is on the proper application of sequential methods, we view the development of
continuous-time methods as an important complementary direction which becomes
especially important in settings with insufficient domain knowledge to make
discretization decisions.
% or when the system is more accurately modeled by a
% continuous time process.

% In general, changing the sequence length, either by changing the outcome
% variable as in \cite{tran2019double} and \cite{gottesman2018evaluating} or by
% changing the discritization bin width as in this work and
% \cite{sofrygin2019targeted}, may change both the estimand and the number of
% samples that follow the treatment regime of interest. As our goal is to isolate
% the impact that sequence length has on the estimators \textit{for fixed
% estimand and sample size}, we focus on estimands that are not impacted by
% changes to the discretization bin width. Specifically, in both our simulated
% and real data analyses, we consider static binary treatment sequences that
% start at zero and may jump to one at some point in the sequence. This type of
% treatment variable can be, and commonly is, used to represent the initiation of
% treatment (e.g. \cite{tran2019double,petersen2014targeted} and others). If
% aligned to the discretization bins, estimands corresponding to treatment rules
% such as ``never treat'', ``treat immediately'', and ``do not treat before time
% $t$'' are unaffected by discretization decisions and we use such rules to
% examine the effects of sequence length and discretization for a fixed estimand.

\section{Motivating application: measuring the effect of early antibiotics for patients with sepsis}

This work is motivated by recent studies that use electronic health records to
estimate the cumulative impact of an exposure or treatment over time, while
adjusting for confounders, e.g., \citep{neugebauer2014targeted, soleimani2017treatment, schulam2017reliable,
komorowski2018artificial, lim2018forecasting, gottesman2018evaluating,sofrygin2019targeted}. In particular, we are motivated
by the study of how early treatment affects patients hospitalized with a
life-threatening complication of infection called sepsis. Sepsis is characterized by organ
failure caused by infection and is one of the leading causes of in-hospital
mortality in the United States \citep{rhee2019prevalence}. Several observational
studies have suggested that early initiation of antibiotic therapy improves
outcomes for patients with sepsis \citep{ferrer2014empiric, liu2017timing, peltan2019ed}; however, these studies have treated the decision of when to initiate antibiotics as a single time-point
intervention and do not account for time-varying confounding. In contrast, we
treat the decision to initiate antibiotics as a time-varying treatment and
estimate the effect of delayed versus early antibiotics
while adjusting for potential time-varying confounding and right censoring.

We used EHR data from 9,523 patients admitted to the
emergency department (ED) with sepsis to estimate the population average
treatment effect (ATE) of receiving antibiotics within one hour of admission
versus not receiving antibiotics during the first 13 hours after admission. (In the latter regime, antibiotics may be given after the first 13 hours.) As sepsis is primarily characterized by organ
dysfunction, our outcome was an aggregate measure of cumulative organ
dysfunction measured 73 hours after admission. As in our simulated
data, we compare the performance of the estimators listed above under different
discretizations of the EHR data. 

\section{Problem definition}
\label{sec:longitudinal_ci}
\subsection{Time discretization and notation}
We consider how changing the temporal discretization of our data affects discrete-time estimators. We do this by first defining a ``finest discretization" of time and then considering various coarser discretizations created by taking a subset of time points from the finest discretization. 
The finest discretization consists of 
a sequence of $T^*$ equally-spaced measurements with $\delta^*$ time between successive measurements. We index these measurements by $t=1,2,\dots,T^*$. For example, if time is measured in hours, then the first measurement time ($t=1$) is $0$ hours after baseline, the second ($t=2$) is  $\delta^*$ hours after baseline,..., and the last ($t=T^*$) is $(T^*-1)\delta^*$ hours after baseline.
This finest discretization (also called the ``uncoarsened discretization") is the gold standard under which our estimand of interest is defined. 
% We also consider coarser discretizations where the time interval between successive measurements  is $\delta > \delta^*$.  
 
Using the finest discretization, 
 consider a sequence of binary-valued treatment variables $\bar{A}(\delta^*) = [A(\delta^*)_1,...,A(\delta^*)_{T^*}]$. %with $\delta^*$ time between the measurement of each variable and sequence length $T^*$.
We are interested in the expected value of an outcome variable $Y$ under a hypothetical treatment regime where everyone in the target  population is set to follow a particular static treatment sequence
$\bar{a}(\delta^*)=[a(\delta^*)_1,\dots,a(\delta^*)_{T^*}]$ \citep{robins1986new,hernan2010causal}.
Using potential outcomes notation, this can be written as $\mathbb{E}[Y^{\bar{a}(\delta^*)}]$, where $Y^{\bar{a}(\delta^*)}$ is the (counterfactual) value of $Y$ had the treatment sequence been set to $\bar{a}(\delta^*)$. 
% Here, $\bar{a}$ may be a fixed sequence of
% treatment values (static policy) or it may depend on other observed features
% (dynamic policy). 

We would like to estimate $\mathbb{E}[Y^{\bar{a}(\delta^*)}]$ using observational data
that has been temporally
discretized with 
$\delta \geq \delta^*$ time between successive time points, where $\delta$ is a multiple of $\delta^*$. Because $\delta$ is a multiple of $\delta^*$, the sequence with bin width $\delta$ contains a \textit{subset} of time points from the finest discretization. The resulting coarsened sequence has length $T(\delta) = \lfloor\frac{\delta^*}{\delta}(T^*-1)\rfloor+1$. 
We denote the dataset discretized at bin width $\delta$ by 
$\mathcal{D}(\delta) = \{(V_i,\bar{A}(\delta)_i,\bar{L}(\delta)_i,Y_i)\}_{i=1}^N$, where for each $i$, $V_i$
represents baseline features, $\bar{A}(\delta)_i = [A(\delta)_{i,1},...,A(\delta)_{i,T(\delta)}]$ represents a
sequence of treatment values, $\bar{L}(\delta)_i = [L(\delta)_{i,1},...,L(\delta)_{i,T(\delta)}]$ represents
a sequence of time-varying features, and $Y_i$ represents the outcome of
interest. 
We sometimes suppress the dependence of sequence length $T(\delta)$, treatment sequence $\bar{A}(\delta)$, feature sequence $\bar{L}(\delta)$, and dataset $\mathcal{D}(\delta)$ on $\delta$.
The temporal ordering of variables is assumed to be: $V_i,L_{i,1},A_{i,1},...,L_{i,T},A_{i,T},Y_i$. We assume that the vectors $(V_i,\bar{A}_i,\bar{L}_i,Y_i), i=1,\dots,N$ are independent, identically distributed draws from an unknown joint distribution $P$.
Where it does not cause ambiguity, we drop the index $i$. 

Denote by $\bar{L}_t = [L_{1},...,L_{t}]$ the feature sequence up to and including time point $t$, $\underbar{L}_t = [L_{t},...,L_{T}]$ the feature sequence from time point $t$ onward, and $\bar{L}_{j:k} = [L_j,...,L_k]$ the feature sequence from time point $j$ through time point $k$, with analogous definitions for $\bar{A}_t$, $\underbar{A}_t$, and $\bar{A}_{j:k}$. Additionally, we will occasionally refer to the potential outcome of $L_t$ under treatment sequence $\bar{a}$ denoted by $L_t^{\bar{a}}$, as well as the potential outcome of $L_t$ when only the treatment subsequence $\bar{a}_{j:k}$ is set, denoted by $L_t^{\bar{a}_{j:k}}$. 
% It is assumed that, for all $t$, $L_{t}$ is measured before $A_{t}$ and that treatment at time $t$ may only impact future measurements (i.e. $L_{t+1},\dots$). That is, the temporal ordering of variables is assumed to be: $V,L_1,A_1,...,L_T,A_T,Y$.
% Let
% $A_t$ denote treatment variable at time $t$, $\bar{A}$ denote the full
% sequence of treatment variables, and $\bar{A}_t$ denote the treatment sequence
% up to and including time $t$ with analogous definitions for $L$. 
% We consider binary treatment sequences that follow a counting
% process (i.e. once the treatment variable jumps to 1, it stays 1). This type of
% treatment variable is frequently used to represent the initiation of a
% treatment or intervention (e.g. \cite{}).

We address the following question: \emph{how does the discretization bin width $\delta$ --- and, by extension, sequence length $T(\delta)$ --- affect various
estimators of $\psi = \mathbb{E}[Y^{\bar{a}(\delta^*)}]$?} We address this question by starting with the finest
discretization (small $\delta$ and large $T(\delta)$) and gradually moving to a coarse
one (large $\delta$ and small $T(\delta)$). In general, changing $T(\delta)$, either by changing the outcome variable as in
\citet{gottesman2018evaluating} and \citet{tran2019double}, or by changing
$\delta$ as in \citet{sofrygin2019targeted}, may change both the
estimand and the number of participants that follow $\bar{a}$. As our goal is to
isolate the impact that $\delta$ has on an estimator \textit{for a fixed estimand and without changing the number of participants who follow $\bar{a}$}, we focus on estimands that are not impacted by
changing $\delta$. We assume that both the observed treatment sequence $\bar{A}$ (with probability 1) and the treatment regime $\bar{a}$ are binary sequences that
start at 0, may jump to 1 at some point, and remain at 1 thereafter; we assume that this holds under the finest discretization, which implies that it holds under any coarser discretization. This type of
restricted treatment sequence can be used to represent the initiation of
a treatment at a given time --- i.e., the jump time.
% As described below, we assume that there exists a sufficiently fine discretization
% that satisfies the necessary causal identifiability assumptions. 

Throughout, we consider static treatment regimes  $\bar{a}(\delta^*)$ that  represent the rules ``treat immediately'', ``never treat",  and ``do not treat before time $k\delta^*$''.
The rule ``treat immediately'' is represented by the sequence of treatments $[1,\dots,1]$, the rule ``never treat'' is represented by the  sequence of treatments $[0,\dots,0]$, and the rule ``do not treat before time $k\delta^*$'' is represented by the sequence of treatments consisting of $0$'s through time $k\delta^*$ and subsequent treatments unspecified. 
For each of these regimes, it follows that the estimand 
 $\mathbb{E}[Y^{\bar{a}(\delta^*)}]$ (which is defined using the finest discretization) equals the corresponding quantity $\mathbb{E}[Y^{\bar{a}(\delta)}]$ (defined in terms of the coarser discretization $\delta>\delta*$)
  as long as  the coarser  discretization includes the measurement time  $0$, $(T^*-1)\delta^*$, or  $k\delta^*$, respectively. 

\subsection{Identifiability assumptions}
\label{sec:identifiability}

All quantities in this paragraph are with respect to the finest discretization, i.e., $\delta^*$, which we suppress in the notation for clarity.
In order for our estimand $\mathbb{E}[Y^{\bar{a}}]$ to be identifiable from the observed data under the finest discretization, we make three assumptions about the data generating distributions. Discussion of these assumptions and their definitions can be found in \citet{hernan2010causal}. 
First, we make the consistency assumption that $Y^{\bar{a}} = Y$ if $\bar{A} = \bar{a}$,  and $\bar{L}_t^{\bar{a}} = \bar{L}_t$ if $\bar{A}_{t-1} = \bar{a}_{t-1}$.
Second, we assume that
there is a positive  probability of $A_t$ jumping to $1$ or remaining $0$ given the history up to time $t$  (positivity) or formally, $0 < P(A_t = 1 \mid A_{t-1} = 0, \bar{L}_{t}, V) < 1$ almost surely, for all $t$,
where $A_{0}$ and $L_0$ are defined as 0 by convention. Finally, we
assume sequential exchangeability
which can be written as $(Y^{\bar{a}},\underbar{L}_{t+1}^{\bar{a}}) \indep A_t \mid \bar{A}_{t-1}=\bar{a}_{t-1},\bar{L}_{t},V$ for all $t$.
%
% where $L_t^{\bar{a}}$ is the potential outcome of $L_t$ under the treatment sequence $\bar{a}$. 
% In certain cases, such as when $\bar{a}$ does not depend on $\bar{L}$ and $Y$ depends on $\bar{L}$ only through $\bar{A}$, a weaker version of sequential exchangeability may be used. We will assume this stronger version, sometimes referred to as \textit{dynamic sequential exchangeability} \cite{hernan2010causal} as it allows for dynamic treatment rules which depend on $\bar{L}$. 
If these assumptions hold, then $\mathbb{E}[Y^{\bar{a}}]$ is identifiable from the observed data. We  make these assumptions only for the finest discretization.  Importantly, assumptions such as sequential exchangeability may fail to hold under a coarser discretization $\delta > \delta^*$. One of our goals is to examine   estimator bias caused by using such a coarser discretization.

\section{Estimators}
\label{sec:estimation_methods}
We compare three estimators that we consider representative of
common approaches to causal estimation from longitudinal data, namely: inverse
probability weighting, an iterative regressions approach, and a targeted minimum loss-based estimator. We do not consider g-estimation which may be used to estimate
the parameters of a structural nested mean model \citep{robins2004optimal}. We
review the details of the estimators used in this work. Note that all three estimators depend implicitly on the bin width $\delta$, but we omit this dependence for conciseness.

\subsection{Inverse probability of treatment weighting (IPW)} 
\label{sec:ipw} 
%
%Based on the principal of importance sampling,  
IPW methods estimate expectations
under a target treatment regime using the outcome data from participants whose observed treatment sequences are consistent with that regime \citep{horvitz1952generalization,robins2000marginal,precup2000eligibility,thomas2016data}. 
% These methods are practically attractive
% because they are easily translated from point treatment to longitudinal
% settings by considering the probability of a treatment sequence rather
% than a single time-point treatment. 
We use a longitudinal IPW estimator with the
following form:
\begin{align}
    \hat{\psi}_{IPW} = \sum_i \frac{\mathbb{I}[\bar{A}_i = \bar{a}]}{\hat{g}(\bar{a};\bar{L}_i,V_i)} Y_{i} \left/ \sum_i \frac{\mathbb{I}[\bar{A}_i = \bar{a}]}{\hat{g}(\bar{a};\bar{L}_i,V_i)}\right.,
\end{align}
where $\hat{g}(\bar{a};\bar{L},V) = \prod_t
\hat{g}_t(a_t;\bar{a}_{t-1},\bar{L}_{t},V)$ and $\hat{g}_t(a_t;\bar{a}_{t-1},\bar{L}_{t},V)$ is an estimate of the probability of $A_t = a_t$ given the features $\bar{L}_t$ and treatment sequence $\bar{A}_{t-1} = \bar{a}_{t-1}$, and where $\mathbb{I}[X]$ is the indicator variable taking value $1$ if $X$ is true and $0$ otherwise.
% IPW  is consistent when the estimate of $g$ is  consistent. IPW estimators may have high variance in longitudinal settings \citep{glynn1994importance}.
% One simple method to reduce this variance, sometimes referred to as weighted
% importance sampling \citep{jiang2016doubly}, is to include a multiplicative control variate which
% results in an estimator which may have lower variance. We will use
% the following weighted IPW estimator throughout the rest of the text:
% %
% \begin{align}
%     \hat{\psi}_{IPW} = \sum_i \frac{\mathbb{I}[\bar{A}_i = \bar{a}]}{\hat{g}(\bar{a};\bar{L}_i,V_i)} Y_{i} \left/ \sum_i \frac{\mathbb{I}[\bar{A}_i = \bar{a}]}{\hat{g}(\bar{a};\bar{L}_i,V_i)}\right. .
% \end{align} 
% %

\subsection{Iterative regression (IR)}
\label{sec:ir}
An alternative approach is based on a direct application of the g-computation formula \citep{robins1986new}. Define $Q_{T} =
\mathbb{E}[Y|\bar{A}_{T}=\bar{a}_{T},\bar{L}_{T},V]$ and, for each $t=T-1,\dots,1$, define $Q_{t} = \mathbb{E}[Q_{t+1}|\bar{A}_{t}=\bar{a}_{t},\bar{L}_{t},V]$. It follows that, if the causal identifiability assumptions hold for bin width $\delta$,
$\mathbb{E}[Y^{\bar{a}(\delta)}] = \mathbb{E}[Q_1]$ \citep{robins1986new,hernan2010causal}. Starting with $Q_{T}$ and working backwards in time, the IR method iteratively
estimates each $Q_t$ using data from subjects who followed $\bar{a}$ through time $t$. A complete description of this algorithm is provided in Section 1.1 of the supplementary materials.
% by regressing the previous estimates (i.e.
% $\hat{Q}_{i,t+1}$) onto the observed covariates up to time $t-1$, using data
% from participants who followed $\bar{a}$ up to time $t-1$. This process is iterated
% until $t=1$, producing a final estimate $\hat{\psi}_{IR}$.
% This method requires specifying a model for each $Q_t$ and the estimator is
% consistent if each of these models is correctly specified.
The consistency of this method generally requires that the model for $Q_t$ at each iteration is correctly specified. 

\subsection{Targeted minimum loss-based estimation (TMLE)} 
\label{sec:tmle}

As stated above, the IPW and IR methods are only consistent so long as $g$ and $Q$, respectively, are consistently estimated. On the other
hand, so called doubly robust estimation methods, of which several exist for
the longitudinal setting, may retain consistency so long as
\textit{either} $g$ or $Q$ can be consistently estimated. In this paper, we do not
examine the behavior of all such doubly robust methods. Instead, we focus on a
single targeted minimum loss-based estimator \citep{van2012targeted} which was shown by \citet{tran2019double} --- in which it was referred to as ``DRICE weighted'' --- to perform well on
sequences up to length six. This TMLE method, introduced by \citet{van2012targeted} and based on methods in \citet{Robins92}, \citet{Robins2000TMLE}, \citet{BangRobins2005TMLE}, and \citet{van2006tmle}, closely resembles the
IR method. However, at each iteration $t$ of the algorithm, TMLE performs a \textit{targeting} step that updates the initial estimate of $Q_t$ by combining it with the estimate of $g$. A full description of the TMLE used in this work is provided in Section 1.2 of the supplementary materials and a similar description can be found in \citet{tran2019double}.
%
%
% As before, the TMLE method begins by regressing $Y$ onto
% $\bar{L}$ among participants who followed $\bar{a}$ up to time $T-1$ to form an
% initial (untargeted) estimate of $Q_{T+1}$. Call this initial estimate
% $\hat{Q}_{T+1}$. Next, using a logistic regression model, $Y$ is regressed onto
% a constant with $\text{logit}\,\, \hat{Q}_{T+1}$ included as a fixed offset and
% using $\frac{\mathbb{I}[\bar{A} = \bar{a}]}{\hat{g}(\bar{a};\bar{L},V)}$ as
% observation weights resulting in a updated (targeted) estimate
% $\hat{Q}^*_{T+1}$. This process is iterated as before to produce a final
% targeted estimate $\hat{\psi}_{TMLE}$.
%Below, we compare thebias and variance of these estimators using simulated data.

\section{The effect of discretization bin width \texorpdfstring{$\delta$}{delta} in simulated data}
\label{sec:synthetic_experiments}
To evaluate the impact of $\delta$ on estimator bias and
variance, we simulated longitudinal data from the finest discretization $\delta^* = 1$ with sequence length $T^* = 257$, and then we constructed coarsened versions of the data at increasing values of $\delta$.
% and smaller corresponding sequence lengths $T(\delta)$. 
We considered a binary
treatment sequence and continuous outcome and we included two baseline
features and three time-varying features. 
% Our data generating mechanism
% follows a Markov process, which, for $\delta = \delta^*$, naturally satisfies the identifiability
% assumptions in Section \ref{sec:identifiability}. 
Specifically, we generated $N
= 1000$ sequences at the finest discretization
% , each with $\delta^* = 1$ and length $T^*=257$, 
according to the 
functional causal model below in which all quantities are with respect to $\delta^*$, which we omit for clarity. In this model, $expit(x) = \frac{1}{1+e^{-x}}$, $V = (V_{1},V_{2})$ are continuous
baseline features, $L_t = (L_{t,1},L_{t,2},L_{t,3})$ are continuous
time-varying features and $L_{t,1:2} = (L_{t,1},L_{t,2})$, $A_t$ is the binary time-varying treatment, $Y$ is
the continuous outcome, and $\boldsymbol{\beta}$ and $\boldsymbol{\gamma}$ are parameters that we set as described below:
\begin{align*}
    V_d &\sim N(0,1),\,\, d = 1,2\\
    L_{t,1} &= \beta_{1,1} + \boldsymbol{\beta}_{1,2}V + \boldsymbol{\beta}_{1,3}L_{t-1} + \beta_{1,4}A_{t-1} + \epsilon_{L,1}\\
    L_{t,2} &= \beta_{2,1} + \boldsymbol{\beta}_{2,2}V + \boldsymbol{\beta}_{2,3}L_{t-1} + \beta_{2,4}A_{t-1} + \epsilon_{L,2}\\
    L_{t,3} &= \beta_{3,1} + \boldsymbol{\beta}_{3,2}V + L_{t-1,3} + \boldsymbol{\beta}_{3,3}L_{t-1,1:2} - \beta_{3,1}A_{t-1} + \epsilon_{L,3}\\
    A_t \mid A_{t-1} = 0 &\sim Bern(expit(\gamma_{1} + \boldsymbol{\gamma}_{2}V + \boldsymbol{\gamma}_{3}L_{t}))\\
	Y &= \beta_{3,1} + \boldsymbol{\beta}_{3,2}V +  L_{T^*,3} + \boldsymbol{\beta}_{3,3}L_{T^*,1:2} - \beta_{3,1}A_{T^*} + \epsilon_{Y}.
\end{align*}

Each error term $\epsilon$ was sampled independently from a $N(0,(0.05)^2)$ distribution that we assume (since this is a functional causal model) not to be impacted by intervening on the treatment sequence $\bar{A}$. The parameter $\gamma_{1}$ was set to $-5.5$ so that approximately $25\%$ of participants follow the ``never treat'' regime. The parameters $\gamma_{3,3}$ and $\beta_{3,1}$ were set to $0.5$ and $0.006$, respectively, to ensure time-dependent confounding by $L_{t,3}$. The remaining parameters were randomly generated independently from a $N(0,(0.005)^2)$ distribution, which was chosen so that most parameters were relatively small, thereby controlling the variance of features at late time steps. Randomly generated parameters were generated once and held constant across all simulations. 

% The treatment value $A_t=1$ can be interpreted as treatment having started
% at or prior to time $t$ so $P(A_t = 1 \mid A_{t-1} = 1) = 1$ (hence the conditioning bar in the expression for generating $A_t$). 
In the above functional causal model, the outcome $Y$ can be thought of as $L_{T^*+1,3}$.
% and $\mathbb{E}\left[L_{t,3}^{a_{t-1}=1} - L_{t,3}^{a_{t-1}=0}\right] =
% -\beta_{3,0}$. That is, the effect of intervention initiation is to decrease the slope of
% $\mathbb{E}[L_{t,3}]$ over time by $\beta_{3,0}$. 
Our target estimand was the expectation of $Y$ under the ``never treat'' regime (i.e.,
$\mathbb{E}[Y^{\bar{a}(\delta^*)=0}]$). 
% We emphasize that, by construction, we do not change the estimand when different bin widths are considered. 
Note that the above functional causal model follows a Markov process, which, for $\delta = \delta^*$, naturally satisfies the causal identifiability assumptions. These assumptions may not hold for coarsened data with bin width $\delta > \delta^*$. 
% We aim to estimate $\mathbb{E}[Y^{\bar{a}=0}]$ based on the uncoarsened data and also based on coarsened data.

To evaluate the performance of the estimation algorithms under different
discretization bin widths, we coarsened the original sequences using bin widths of $\delta = 1,2,4,8,...,256$, where $\delta = \delta^* = 1$ results in the uncoarsened data. As discussed above, all values of $\delta$ are multiples of $\delta^* = 1$, and thus this coarsening can be thought of as taking a subset of time points from the finest discretization. 
% Specifically, for bin width $\delta$, the coarsened data includes the discrete indices $s(\delta) = \{1,1+\delta,1+2\delta,...,1 + \delta \lfloor (T^*-1)/\delta\rfloor\}$. 
% By choosing $\delta$ to be powers of $2$, we ensured that the first and last time-steps from the uncoarsened sequence were always included.
As before, let $T(\delta)$ be the length of the sequence with bin width $\delta$. This coarsening scheme results
in sequence lengths ranging from $T(1) = T^* = 257$ to $T(256) = 2$ time points. 
% We adopted the measurement ordering convention from \citet{sofrygin2019targeted} whereby if treatment changes from 0 to 1 within a discretization window, the covariate values from the time of the change are carried forward to the end of the window (for details on this convention and various potential edge cases, see \citep{msmstruct}).
% coarsened the
% sequences using the same discretization procedure used in
% \citet{sofrygin2019targeted}. This procedure uses ``last value carried
% forward'' with the exception that the temporal ordering between
% covariates and changes in treatment is maintained. This means that if treatment changed from a 0 to a 1 within a discretization bin, the covariate values from just before the change would be carried forward. 
An important property of this
data generating mechanism and coarsening procedure is that the last ($t = T^* = 257$)
time point of the finest discretization is always retained. As described above, for the restricted binary treatment sequences we consider, this means that $\mathbb{E}[Y^{\bar{a}(\delta^*)=0}] = \mathbb{E}[Y^{\bar{a}(\delta)=0}]$ under all settings of $\delta$. The
number of samples that follow $\bar{a}(\delta)$ also remains the same for all $\delta$.
% That is, we are varying the sequence length while keeping the number
% of samples fixed.

For each bin width $\delta$ and estimator $\hat{\psi}$, we estimated the absolute bias $|\mathbb{E}[Y^{\bar{a}(\delta^*)}] - \mathbb{E}[\hat{\psi}(\delta)]|$, variance $Var(\hat{\psi}(\delta))$, and mean squared
error (MSE) $\mathbb{E}[(\mathbb{E}[Y^{\bar{a}(\delta^*)}] - \hat{\psi}(\delta))^2]$ using 1000 datasets of size $N=1000$ generated using this
data generating mechanism and coarsened according to $\delta$. For IPW and TMLE, $g_t  = P(A_t = 1 \mid A_{t-1} = 0, \bar{L}_{t},V)$ was estimated for each $t$ using the parametric model $m_t(\bar{L}_{t},V;\boldsymbol{\theta}_t) = expit(\theta_{t,0} + \boldsymbol{\theta}_{t,1} V + \boldsymbol{\theta}_{t,2} L_{t})$. The parameter vector $\boldsymbol{\theta}_t$ was estimated using data from all participants $i$ for whom $A_{i,t-1} = 0$. Additionally, for IR, $Q_t = \mathbb{E}[Q_{t+1} \mid \bar{A}_{t} = \bar{a}_{t},\bar{L}_{t},V]$ was estimated for each $t$ using the parametric model $q_t(\bar{L}_{t},V;\boldsymbol{\nu}_t) = \nu_{t,0} + \boldsymbol{\nu}_{t,1}L_t + \boldsymbol{\nu}_{t,2}V$.
As described in Section \ref{sec:ir}, the parameter vector $\boldsymbol{\nu}_t$ was estimated using data from participants who followed $\bar{a}$ through time $t$. For TMLE, this model was used for the initial estimate of $Q_t$ before the additional targeting step described in Section \ref{sec:tmle}.
% \subsection{Randomized treatment}
%
% We first estimated the MSE, bias, and variance of all three estimators when
% $A_t$ is sampled without confounding (i.e. $\beta_{a1} = \beta_{a2} =
% \mathbf{0}$). This emulates a randomized trial where subjects are assigned a
% random treatment start time. Results are shown in Figure \ref{fig:zero_effect}.
% As expected, we observed that variance increased with sequence length for all
% estimators with a sharp increase in variance for long sequences. Importantly,
% this increase is most severe in the IPW estimator, followed by TMLE and then IR
% which shows little change in variance as a function of sequence length. In
% addition, we observed that bias for the IR and TMLE methods increased
% substantially as the resolution length decreased, despite the lack of
% confounding.
%
% \begin{figure*}[t!]
%     \centering
%     \begin{subfigure}[b]{0.9\textwidth}
% 		\includegraphics[width=\textwidth]{figures/mse_continuous_rct_True_zero_effect_False_IR_TMLE_IPW.pdf}
% 	\end{subfigure}\\
%     \begin{subfigure}[b]{0.9\textwidth}
% 		\includegraphics[width=\textwidth]{figures/bias_continuous_rct_True_zero_effect_False_IR_TMLE_IPW.pdf}
%     \end{subfigure}\\
%     \begin{subfigure}[b]{0.9\textwidth}
% 		\includegraphics[width=\textwidth]{figures/variance_continuous_rct_True_zero_effect_False_IR_TMLE_IPW.pdf}
% 	\end{subfigure}\\
%     \caption{}\label{fig:zero_effect}
% \end{figure*}
%
% \subsection{Confounded treatment}
% \label{sec:confounded_synth}
\subsection{Results}

\ifinlinefigs
\begin{figure}[t!]
    \input{sections/fig1}
\end{figure}
\fi

The bias, variance, and MSE for each estimator under different $\delta$ are shown in Figure
\ref{fig:non_zero_effect}. Though somewhat hard to see in the Figure, variance increases
with sequence length for all estimators with a sharp increase in variance for
the TMLE and IPW estimators on long sequences. This increase is most severe for the IPW estimator,
followed by TMLE and then IR, the last of which does not show the same steep increase in variance as the other two. In Section 3 of the supplementary material, we review and evaluate several approaches to reducing this variance.

Additionally, we observe two distinct forms of bias. First, the IPW estimator exhibits increased bias in the
longest sequences. This is likely due to finite sample bias caused by near-positivity violations. Specifically, as $\delta$
becomes smaller, the probability of a participant beginning treatment within each
time bin grows smaller until, for small enough $\delta$, we observe bins containing no
jumps. Consistent with previous studies
\citep{petersen2012diagnosing,tran2019double}, we observe that these finite
sample positivity violations have a much smaller effect on the IR and
TMLE methods than the IPW method. In the presence of positivity
violations the IR an TMLE methods must extrapolate when estimating $Q$. If the
model for $Q$ is a correctly specified parametric model (as it is in this example), this
extrapolation may result in relatively little bias; however, due to potential model misspecification, we cannot expect this in general. 

We observe a second form of bias when $\delta$ is large and $T(\delta)$ is small. Importantly, for all estimators, this bias occurs
\emph{even when there is no confounding} (see Section 2 of the supplementary material for an
example). 
%This type of finite sample positivity violation
% may be addressed by pooling across time.
This second form of bias dominates MSE for large $\delta$, whereas variance dominates MSE for small $\delta$. In the next section, we discuss this second source of bias in the IR and TMLE methods and give recommendations for avoiding it.

\section{Discretization bias in IR and TMLE}
\label{sec:discretization_bias_ir}
% As we demonstrated in the previous section, choosing how to discretize one's
% data represents a bias/variance trade-off. The specifics of this trade-off will
% vary depending on the application and, in this section, we provide intuition
% for the source of this bias and how to choose an appropriate discretization bin
% width. 
\subsection{Probability limits of estimators using uncoarsened vs. coarsened data}
Suppose that we have discretized our data using bin width $\delta^*$ and, as in the previous section, we coarsen the finest discretization by using a subset of its time points. We show that if, in the coarsened sequence, changing a treatment value within a discrete time bin can have a causal effect on feature values within the same time bin,  then the IR and TMLE estimators applied to the coarsened data may result in inconsistent estimators of $\mathbb{E}[Y^{\bar{a}(\delta^*)}]$ even under correct model specification and without time-dependent confounding. We further show that this result explains the bias observed in Section \ref{sec:synthetic_experiments} when using IR and TMLE with large $\delta$.

Throughout the remainder of this section, all quantities are with respect to the finest discretization, $\delta^*$, unless otherwise specified. 
As in our simulated data experiments, we will assume that the finest discretization satisfies a Markov property with respect to
$\bar{L}$ and Y, that is, $L_t \indep \bar{L}_{t-2} \mid \bar{A}_{t-1},L_{t-1}$ and $Y \indep \bar{L}_{T^*-1} \mid \bar{A}_{T^*},L_{T^*}$. We will assume that the finest discretization obeys the causal identifiability assumptions, but we will not assume that the coarsened sequence does. For simplicity, we assume that all baseline features are included in $L_1$ (so there is no $V$).

As before, consider coarsening this sequence to have bin width $\delta$, where $\delta$ is a multiple of $\delta^*$. Denote by $\bar{s} = \{s(1),...,s(T(\delta))\} \subset \{1,...,T^*\}$ the subset of indices from the finest discretization included in the coarsened one where $s(k)$ is the $k$'th value in $\bar{s}$. By definition, $\bar{s}$ has length
$T(\delta)$ and we assume that the first and last indices of the uncoarsened sequence are included in $\bar{s}$, i.e., 
$s(1) = 1$ and $s(T) = T^*$. It follows that $L_{s(k)} = L(\delta)_k$ and the same holds replacing $L$ by $A$.
% Typically, $\bar{s}$ will consist of
% evenly spaced indices, however, this need not be the case in general. 
% For parsimony, let
% $\bar{A}_{\bar{s}} = [A_{s(1)},A_{s(2)},...,A_{s(T)}] = \bar{A}(\delta)$ be the coarsened treatment sequence, with analogous definitions for
% $\bar{L}_{\bar{s}}$, $\bar{l}_{\bar{s}}$, and $\bar{a}_{\bar{s}}$.
Our interest is in the difference between the probability limits of a given estimator when applied to the finest discretization and to the coarsened discretization  $\bar{s}$; we separately  determine this for the IR and TMLE estimators. Recall that, for any  discretization of the data, the IR  estimator converges in probability to the g-computation formula applied according to that discretization, as long as the corresponding $Q$ is consistently estimated; this holds as well for the TMLE if either the corresponding $g$ or $Q$ is consistently estimated. We show how the coarsened and uncoarsened g-computation formulas differ and describe conditions under which the two are equal.

% We begin with an example illustrating how the coarsened and uncoarsened g-computation formulas can differ on data with no time-dependent confounding. Consider the following simple functional causal model for $t = 1,2,3$:
% %
% \begin{align}
%     A_t \mid A_{t-1} = 0 &\sim Bern(\pi)\\
%     L_t &= L_{t-1} + \eta - A_{t-1}\\
%     Y &= L_3 + \eta - A_3,
% \end{align}
% where $L_0 = A_0 = 0$, $\pi \in (0,1)$, and $\eta\in\mathbb{R}$. Note that $\bar{A}$ is fully randomized and, thus, is unconfounded with $Y$. In a functional causal model, we can plug in the values $\bar{A}=0$ to calculate the expected potential outcome $\mathbb{E}[Y^0] = 4\eta$. Now consider applying the g-computation formula to data coarsened according to $\bar{s} = \{1,3\}$ (i.e., dropping $t=2$) using the treatment regime $\bar{a} = 0$. Using the definition of $Q_t$ from Section \ref{sec:ir}, we find that applying the g-computation formula to the coarsened data results in a different value than $\mathbb{E}[Y^{0}]$:
% %
% \begin{align}
%     Q_2 &= \mathbb{E}[Y \mid \bar{L}_{s(2)},\bar{A}_{s(2)} = 0] = L_3 + \eta\\
%     Q_1 &= \mathbb{E}[Q_2 \mid \bar{L}_{s(1)},\bar{A}_{s(1)} = 0] = L_1 + 3\eta - \pi\\
%     \mathbb{E}[Q_1] &= 4\eta - \pi \neq \mathbb{E}[Y^0].
% \end{align}

% We now consider the g-computation formula for general data and $\bar{s}$. 
For conciseness of notation, we write expressions such as $P(X=x\mid Z = z)$ and $\mathbb{E}[X \mid Z = z]$ as $P(X=x \mid z)$ and $\mathbb{E}[X \mid z]$, respectively. 
E.g., we write $P(L_{k} = l_{k} \mid \bar{A}_{k-1}= \bar{a}_{k-1}, L_{k-1}=l_{k-1})$ as 
$P(L_{k} = l_{k} \mid \bar{a}_{k-1}, l_{k-1})$.
Using the Markov assumption and assuming that each $L_t$ is discrete valued, the  \textit{g-computation formula applied to the coarsened sequence} can be written as
\begin{align}
	\label{eq:disc_gform}
% 	\sum_{\bar{l}_{\bar{s}}} \mathbb{E}[Y \mid \bar{a}_{\bar{s}}, l_{T^*}] \prod_{k=1}^{T} P(L_{s(k)} = l_{s(k)} \mid \bar{a}_{\bar{s}(k-1)}, l_{s(k-1)}),
	\sum_{\bar{l}(\delta)} \mathbb{E}[Y \mid \bar{a}(\delta), l(\delta)_{T(\delta)}] \prod_{k=1}^{T(\delta)} P(L(\delta)_{k} = l(\delta)_{k} \mid \bar{a}(\delta)_{k-1}, l(\delta)_{k-1}),
\end{align}
where the sum over $\bar{l}(\delta)$ indicates a sum over all possible coarsened feature sequences and conditioning on $\bar{a}(\delta)$ indicates conditioning on the coarsened treatment sequence. We now highlight the differences between this quantity and the uncoarsened g-computation formula. Recalling that $\bar{l}_{j:k} = [l_j,l_{j+1},\dots,l_k]$ and that $L_t^{\bar{a}_{j:k}}$ is the potential outcome of $L_t$ when only the treatment subsequence $\bar{a}_{j:k}$ is set, we can rewrite the \textit{uncoarsened g-computation formula} as
\begin{align}
    \label{eq:po_gform_l1}
	\mathbb{E}\left[Y^{\bar{a}}\right] &= \sum_{\bar{l}} \mathbb{E}[Y \mid \bar{a}, l_{T^*}] \prod_{t=1}^{T^*} P(L_{t} = l_{t} \mid \bar{a}_{t-1}, l_{t-1})\\
	%
% 	\label{eq:po_gform_l2}
% 	\begin{split}
% 	&= \sum_{\bar{l}_{\bar{s}}} \sum_{\bar{l}_{s(1)+1:s(2)-1}} \dots \sum_{\bar{l}_{s(T-1)+1:s(T)-1}} \mathbb{E}[Y \mid \bar{a}, l_{T^*}] P(L_1 = l_1)\\ &\mspace{270mu}\cdot \prod_{k=2}^{T} \left\{\prod_{t = s(k-1)+1}^{s(k)} P(L_{t} = l_{t} \mid \bar{a}_{t-1}, l_{t-1})\right\}
% 	\end{split}\\
% 	%
% 	\label{eq:po_gform_l3}
% 	&= \sum_{\bar{l}_{\bar{s}}} \mathbb{E}[Y \mid \bar{a}, l_{T^*}] P(L_1 = l_1) \prod_{k=2}^{T}\,\, \sum_{\bar{l}_{s(k-1)+1:s(k)-1}} \left\{\prod_{t = s(k-1)+1}^{s(k)} P(L_{t} = l_{t} \mid \bar{a}_{t-1}, l_{t-1})\right\}\\
	%
% 	&= \sum_{\bar{l}(\delta)} \mathbb{E}[Y \mid \bar{a}, l_{T^*}] P(L_1 = l_1) \prod_{k=2}^{T} P(L^{\bar{a}_{s(k-1)+1:s(k)-1}}_{s(k)} = l_{s(k)} \mid \bar{a}_{s(k-1)}, l_{s(k-1)}),\\
	\label{eq:po_gform}
	\begin{split}
	&= \sum_{\bar{l}(\delta)} \mathbb{E}[Y \mid \bar{a}, l(\delta)_{T(\delta)}] P(L(\delta)_1 = l(\delta)_1)\\ &\qquad\qquad\qquad\qquad\cdot\prod_{k=2}^{T(\delta)} P(L(\delta)^{\bar{a}_{s(k-1)+1:s(k)-1}}_{k} = l(\delta)_{k} \mid \bar{a}_{s(k-1)}, l(\delta)_{k-1}),
    \end{split}
\end{align}
where the sum over $\bar{l}$ in Equation \ref{eq:po_gform_l1} indicates a sum over all possible values of the uncoarsened feature sequence. A complete derivation of Equation \ref{eq:po_gform} is given in Section 4.1 of the supplementary materials.
% Equation \ref{eq:po_gform_l2} partitions the sum over $\bar{l}$ into the sum over the  components in $\bar{l}_{\bar{s}}$ (i.e., those in the coarsened sequence: $s(1),\dots,s(T)$) and  over the remaining components; Equation \ref{eq:po_gform_l3} follows from factorizing the nested sums in Equation \ref{eq:po_gform_l2} and the assumption that $s(1)=1$ and $s(T)=T^*$; and Equation \ref{eq:po_gform} follows by recognizing that (by the Markov assumption) the inner sum in Equation \ref{eq:po_gform_l3} is an instance of the g-computation formula applied to ``outcome" $\mathbb{I}[L_{s(k)}=l_{s(k)}]$ setting treatments at each time $s(k-1)+1$ through $s(k)-1$ and conditioning on the (uncoarsened) history through $s(k-1)$. 
% We next compare 
% Equations \ref{eq:disc_gform} and \ref{eq:po_gform}, which represent the coarsened and uncoarsened g-computation formulas, respectively. 
The coarsened (Equation \ref{eq:disc_gform}) and uncoarsened (Equation \ref{eq:po_gform}) g-computation formulas differ in two ways. 
First, all expressions in Equation \ref{eq:disc_gform} condition on the coarsened treatment sequence $\bar{a}(\delta)$, whereas expressions in Equation \ref{eq:po_gform} condition on the uncoarsened treatment sequence $\bar{a}$. 
Second, in Equation \ref{eq:disc_gform} the expected outcome $\mathbb{E}[Y \mid \bar{a}(\delta), l(\delta)_{T}]$ is multiplied by the product of terms $P(L(\delta)_{k} = l(\delta)_{k} \mid \bar{a}(\delta)_{k-1}, l(\delta)_{k-1})$, while in Equation \ref{eq:po_gform} the expected outcome is multiplied by the same product but with $L(\delta)_{k}$ replaced by the intervened-on measurement $L(\delta)^{\bar{a}_{s(k-1)+1:s(k)-1}}_{k}$.
It follows that 
\textit{if, for all $k$, (i) conditioning on $\bar{a}(\delta)_k$ is equivalent to conditioning on $\bar{a}_{s(k)}$ and (ii) the sequence of treatments between times $s(k-1)$ and $s(k)$ (exclusive) does not have a causal effect on the subsequent measurement $L(\delta)_{k}$, then the coarsened and uncoarsened g-computation formulas are equal.} Therefore, under conditions (i) and (ii), and under the assumptions necessary to guarantee that IR and TMLE converge in probability to the coarsened g-computation formula --- including consistent estimation of $Q$ and/or $g$ --- as well as the causal identifiability assumptions for the \textit{finest} discretization, IR and TMLE applied to the coarsened data converge in probability to the original estimand $\mathbb{E}[Y^{\bar{a}(\delta^*)}]$. Importantly, conditions (i) and (ii) may be violated even when $\bar{L}$ has \textit{no} causal effect on $\bar{A}$, which means that coarsening the finest discretization may still induce bias, asymptotically, in IR and TMLE when there is no time-dependent confounding.
The  binary treatment sequences that we consider in this work satisfy condition (i) when considering $\bar{a} = 0$ or $\bar{a} = 1$. Additional work is necessary to investigate other types of treatment sequences and settings that do not follow the Markov assumption.

% To gain further intuition for how Equations \ref{eq:disc_gform} and \ref{eq:po_gform} can differ, c
%

\subsection{Discretization bias in simulations}

\ifinlinefigs
\begin{figure}[t!]
    \input{sections/fig2}
\end{figure}
\fi

We return to the simulated data experiments from Section \ref{sec:synthetic_experiments}. Note the  following: the data generating distribution described in Section \ref{sec:synthetic_experiments} satisfies the Markov assumption; for all $\delta$, the ``never treat'' regime we considered satisfies condition (i) above; the parametric models for $Q_t$ are correctly specified for coarsened data; and, the causal identifiability assumptions are satisfied at  $\delta^*$. Thus, the bias we observe when using IR and TMLE with $\delta > \delta^*$ (Figure \ref{fig:non_zero_effect}) may be due to violations of condition (ii). We illustrate this point by modifying the data generating mechanism from
Section \ref{sec:synthetic_experiments} to include an effect delay $\omega$
such that, in the data generating mechanism, $L_t$ now depends on $A_{t-\omega}$ rather than $A_{t-1}$
(by convention, $A_t = 0$ for all $t < 1$). All other aspects of the data generating mechanism, estimation methods, and experimental setup remained the same, including the target estimand, $\mathbb{E}[Y^{\bar{a}(\delta^*)=0}]$. In particular, $\omega = 1$ is exactly equivalent to the experimental setup in Section \ref{sec:synthetic_experiments}. In this new setup, condition (ii) is satisfied for all $\delta \leq \omega$. Figure \ref{fig:effect_delay}
shows how changing the discretization bin width $\delta$ impacts the bias of the IR estimator under different settings of the effect delay, $\omega$. As predicted by the above
analysis, we see that the bias remains near zero as long as $\delta \leq \omega$. For $\delta > \omega$ the bias grows as in Section \ref{sec:synthetic_experiments}. The same plot for the TMLE estimator is very similar and can be found in Section 4.2 of the supplementary materials.
%
% Practically, this suggests that, when deciding on the bin width $\delta$, we need to consider the effects of changes
% in treatment \emph{within} each discretization bin. Discretization decisions
% should be governed, in part, by how quickly we expect to see treatment decisions affect
% our features. We recommend that practitioners choose the largest bin width such that, if treatment were to change within a discrete time bin, they would not
% expect to observe an effect of that change on feature variables within the same bin. 
% In many circumstances, however,
% the widest unbiased bin width may still produce long sequences. 
In the next
section, we evaluate the impact of $\delta$ in the context of a real clinical causal estimation problem: estimating the effect of antibiotics timing in patients with sepsis.

\section{Estimating the effect of delayed versus immediate antibiotics for patients with sepsis}
\label{sec:real_data_experiments}
% \begin{figure*}[t!]
%     \input{sections/fig3}
% \end{figure*}

% In the preceding sections, we used synthetic data to investigate the effects of
% discretization bin width on bias and variance finding a bias/variance
% trade-off. In this section, we repeat this comparison in the context of a real
% clinical decision making problem: estimating the effect of antibiotics timing
% in patients with sepsis. As previously stated, 
Several observational studies
have suggested that delaying the initiation of antibiotics in patients with
sepsis leads to poorer health outcomes; however, these studies do not account
for potential time-dependent confounding or right censoring. In this section, we
use the estimators described in Section \ref{sec:estimation_methods} to examine
the impact of delayed antibiotics on later organ dysfunction, a primary symptom
of sepsis. We estimated the
average treatment effect (ATE) of receiving antibiotics within one hour of admission versus
not receiving antibiotics within the first 13 hours after admission on the number of dysfunctional organ systems 73 hours after admission. Our data included electronic health records from 9,523
patients admitted to the emergency department at one of  three hospitals in the
Johns Hopkins Hospital system between 2016 and 2018. Patients were
included if they were above age 18 and were classified as having sepsis (according to the
definition introduced in \citet{henry2019comparison}) with onset occurring within the first 24
hours after admission. The number of dysfunctional organ systems was measured
as the sum of the nine organ dysfunction criteria defined in
\citet{henry2019comparison}. Summary statistics  are shown in Table \ref{tab:data_stats}. 

Antibiotics initiation was represented as a binary,  time-varying treatment. We adjusted for
potential time-varying confounding and for right-censoring that occurs when
a patient either leaves the hospital against medical advice or is transferred
to another critical care facility. If a patient died or was discharged (excluding transfers to critical care facilities) before 73 hours after admission, their outcome was measured as the number of dysfunctional organ systems at the time of death or discharge and such patients were \textit{not} considered right-censored. If a patient died or was discharged prior to receiving antibiotics, it is assumed that they never subsequently received antibiotics. The complete list of baseline and time-varying features is given in Section 5.1 of the supplementary materials and includes various demographics, comorbidities, lab measurements, and vital signs. All continuous variables were discretized into five quantile bins.
% %
% \begin{enumerate}
% 	\item Positive if the patient was on mechanical ventilation
% 	\item Positive if that patient was on vasopressors
% 	\item Positive if the most recent creatine measurement was $\geq 1.5$
% 	\item Positive if the most recent bilirubin measurement was $\geq 2.0$
% 	\item Positive if the most recent platelet count was $< 100$ and the patient was not suffering from GI bleeding
% 	\item Positive if the most recent GCS measurement was $< 13$ and the patient was not suffering from stroke or drug overdose and was not sedated
% 	\item Positive if the most recent INR measurement was $\geq 1.5$ or the most recent PTT measurement was $> 60$ and the patient was not on anticoagulants
% 	\item Positive if the most recent heart rate measurement was below 90 bpm, the patient's heart rate had fallen by more than 40 bpm in the last 15 minutes, \textbf{or} the last MAP measurement was lower than 65
% 	\item Positive if the most recent serum lactate measurement was $\geq 2.0$ mmol/L
% \end{enumerate}

\ifinlinefigs
\begin{table}[t]
    \center
\begin{tabular}{rccc}
	& Full Sample & Abx. $\leq$ 1 hour & Abx. $>$ 13 hours\\ \hline
	Number of Participants & 9,523 & 624 & 1,177 \\
	Age & 63.30 (17.68) & 65.68 (16.82) & 61.07 (17.25) \\
	Sex=female & 4711 (49.5\%) & 340 (54.5\%) & 597 (50.7\%)\\
	Charlson Comorbidity Index (CCI) & 7.50 (4.01) & 7.45 (3.70) & 7.65 (4.17)\\
	In-hospital mortality & 1126 (11.8\%) & 106 (17.0\%) & 174 (14.8\%)\\
	No. organ dysfunctions at 13hrs & 1.35 (1.31) & 1.82 (1.51) & 1.22 (1.17)\\
	No. organ dysfunctions at 25hrs & 1.26 (1.33) & 1.61 (1.49) & 1.29 (1.27)\\
	No. organ dysfunctions at 49hrs & 1.09 (1.32) & 1.38 (1.51) & 1.26 (1.39)\\
	No. organ dysfunctions at 73hrs & 0.99 (1.31) & 1.23 (1.44) & 1.21 (1.41)\\
	In-hospital mortality w/in 73hrs & 434 (4.6\%) & 63 (10.1\%) & 46 (3.9\%)\\
	Discharged w/in 73hrs & 419 (4.4\%) & 21 (3.4\%) & 21 (1.8\%)\\
	Censored w/in 73hrs & 341 (3.6\%) & 24 (3.8\%) & 22 (1.9\%)\\
\end{tabular}
\caption{Summary statistics for the Full Sample, patients who received antibiotics within an hour of admission (Abx. $\leq$ 1 hour), and patients who had not received antibiotics in the first 13 hours  after admission (Abx. $>$ 13 hour). Binary variables are summarized as the count and (percentage). Continuous variables are summarized as the mean and (standard deviation). Note that the average number of dysfunctional organ systems decreased over time as patients stabilized or recovered. For patients who died or were discharged before time $t$, their number of organ dysfunctions at time $t$ was recorded as the number of organ dysfunctions at the time of death or discharge.}
\label{tab:data_stats}
\end{table}
\fi

% We treated antibiotics initiation as a time-varying treatment and adjusted for
% potential time-varying confounding as well as right-censoring that occurs when
% a patient either leaves the hospital against medical advice or is transferred
% to another critical care facility. We adjusted for potential confounding by
% several baseline (e.g. age, biological sex, comorbidity burden) and
% time-varying (e.g. blood pressure, temperature, white blood cell count)
% variables (for more details on variable definitions and data, see Appendix
% \ref{app:data}).
We choose $\delta^* = 1$ hour as our finest discretization and introduce a binary censoring variable $\bar{C}(\delta^*) = [C(\delta^*)_1,...,C(\delta^*)_{T^*}]$, where $C(\delta^*)_t = 1$ indicates that a patient was censored at or before time $t$. The temporal ordering of variables under the finest discretization is assumed to be
\begin{align*}
    V,L(\delta^*)_1,C(\delta^*)_1,A(\delta^*)_1,...,L(\delta^*)_T,C(\delta^*)_T,A(\delta^*)_T,Y.
\end{align*} 
For the finest discretization, we make the causal identifiability assumptions and assume that they apply to $\bar{C}(\delta^*)$ as well as $\bar{A}(\delta^*)$. Because we are only interested in antibiotics administration before hour 13, we consider static treatment regimes that set the value of treatment through hour 13 and leave treatment after hour 13 unspecified. That is, anyone who does not have antibiotics administered in the first 13 hours (regardless of treatment after hour 13) is following the treatment regime $\bar{a}(\delta^*) = 0$. Likewise, anyone who receives antibiotics in the first hour after admission is following regime $\bar{a}(\delta^*) = 1$. Our target estimand is  $\mathbb{E}[Y^{\bar{a}(\delta^*)=0,\bar{c}(\delta^*) = 0} - Y^{\bar{a}(\delta^*)=1,\bar{c}(\delta^*) = 0}]$. 

We estimated $g$ and $Q$ using modified versions of the parametric models described in Section \ref{sec:synthetic_experiments}. The parametric model for $g_t$ was modified to incorporate the assumption that a patient cannot initiate antibiotics or become newly censored after death or discharge. We used a parametric model of the same form as that used for $g_t$ to estimate the probability of becoming newly censored at time $t$. The parametric model for $Q_t$ was modified to incorporate the assumption that a patient's outcome cannot change after death or discharge. Full details for these models are given in Section 5.2 of the supplementary materials. 95\% confidence
intervals were constructed using percentile-based non-parametric bootstrap using 500 bootstrap replicates.

We compared the performance of the three estimators described in Section
\ref{sec:estimation_methods} using discretization bin widths of $\delta = 1$, $3$, and $6$
hours corresponding to sequences with $T(\delta) = 73$, $25$, and $13$ time points, respectively. These discretizations always included hours $1$, $13$, and $73$ as time points and thus, changing $\delta$ did not change $\mathbb{E}[Y^{\bar{a}(\delta),\bar{c}(\delta)=0}]$ under the treatment regimes considered. We make causal identifiability assumptions \textit{only} for bin width $\delta = \delta^* = 1$ hour.

\subsection{Results} 

\ifinlinefigs
\begin{figure}[t!]
    \input{sections/fig3}
\end{figure}
\fi

Figure \ref{fig:hcgh} shows the
estimated ATE for each of the three estimators using sequences of length $T(\delta) = 13$, $25$, and $73$. Using the finest discretization ($\delta=\delta^* = 1$ hour), all three estimates of the ATE show that the effect of delayed versus immediate antibiotics was an increase in the number of dysfunctional organ systems at 73 hours. Further, all three estimated 95\% confidence intervals exclude zero. In contrast, a naive estimate of the ATE --- estimated by taking the difference in average outcomes in the two treatment groups --- show a near-zero effect ($-0.01$ $[-0.11 \text{ to } 0.12]$). 
% While literature suggests
% that delays in antibiotics of even an hour can have an effect on patients
% \citep{liu2017timing, peltan2019ed, ferrer2014empiric}, there is little
% literature on how quickly antibiotics may have an observable effect on our
% covariates. Clinical guidelines recommend reassessment of
% empirical antibiotics (antibiotics given without exact knowledge of the source of infection) every 24 hours, though it is unclear upon what evidence these recommendations are based and whether they reflect how early the effect of antibiotics is detectable \citep{schmidt2018evaluation}. 
We find that, for each of the three estimators, 
the point estimates are largely consistent across $\delta = 1,3,6$.
% , suggesting that the effects of antibiotics within 6
% hours are small. 
At the finest
discretization, the IR method results in the narrowest confidence interval,
followed by TMLE, and then by IPW. Confidence interval widths range from 0.28 to 0.30 for IR, from 0.39 to 0.43 for TMLE, and from 0.43 
to 0.50 for IPW. In Section 4 of the supplementary material, we evaluate the impact of weight clipping and pooling on these estimates.

%Finally, temporal pooling when estimating $Q$ had the largest impact of any of
%the variance reduction methods reducing the confidence interval width for the
%IR estimator from around 0.20 to around 0.12 without a noticeable change in the
%point estimate.

% \section{Related Work}
% \label{sec:background}
% \input{sections/background}

\section{Discussion}
\label{sec:discussion}
It has become increasingly viable to evaluate hypothetical treatment regimes
using non-experimental data thanks to the availability of high-resolution logged
data. Before applying discrete-time longitudinal causal estimation methods to
these problems, it is necessary to understand how the behavior of
these methods is impacted by temporally discretizing the data. We evaluated the impact of discretization
bin width on three causal estimation methods, inverse
probability weighting (IPW), an iterated regressions method (IR), and a targeted minimum loss-based
estimator (TMLE), when estimating an expected potential outcome under a binary treatment sequence with at most a single 0-to-1 transition. These results highlight two main takeaways for those hoping to apply these methods.

First, results from both simulated and clinical data corroborated previous
 results \citep{jiang2016doubly,thomas2016data} showing that IR has substantially lower variance in long sequences than IPW, with TMLE somewhere between the two.
This latter result is in contrast with a previous comparison of IR and TMLE \citep{tran2019double}; however, we found that the variance gap only
appeared for relatively long sequences whereas \citet{tran2019double}
considered sequences up to length six. This suggests that the double-robustness property of TMLE may come at the cost of increased variance in long sequences; however, the size of the variance gap may depend on the application.
% and, consistent with recommendations given in \citet{hernan2010causal}, practitioners should test the sensitivity of their conclusions to the choice of estimator.
% Consistent with
% recommendations given in \citet{hernan2010causal}, practitioners should test
% the sensitivity of their conclusions to the choice of estimation method.
% This has two implications for
% applications of these methods. First, despite it's ease of implementation, IPW
% has undesirable variance properties in long sequences. In general, practitioners should consider using a doubly robust method,
% such as TMLE, as such methods retain consistency, but have much better
% variance properties in long sequences. With that said, the double robustness property of
% TMLE may come at the cost of increased variance and, consistent with
% recommendations given in \citet{hernan2010causal}, practitioners should test
% the sensitivity of their conclusions to the choice of estimation method.

Second, we found on simulated data that overly-wide discretization bins led to
substantial bias in all three estimators. For IR and TMLE, we
explored the source of this bias and found that it is governed, in part, by whether a change in treatment can effect features within the same discretization bin. Our analysis suggests a rule-of-thumb for choosing
discretization bin widths when using these estimators: Choose the widest
bins such that changing treatment within a bin does
not affect features within the same bin. For example, in our application, if we do not believe the effect of
antibiotics will be noticeable until at least six hours after administration,
this suggests six hours as a plausible bin width. If instead we were to consider
the effects of a fast acting medication, this bin width
would need to be smaller. When such domain knowledge is
unavailable, practitioners should exercise caution when applying discrete-time
methods to continuous-time data and should consider continuous-time causal
estimation methods.
% such as those in 
% \citet{schulam2017reliable}, \citet{soleimani2017treatment}, and \citet{zhang2011causal}. 
We emphasize
this particularly in the case of applications of RL to continuous-time decision
making settings where it is standard practice to model the system as a
discrete-time time-homogeneous Markov decision process, e.g., \citep{jiang2016doubly,thomas2016data,gottesman2018evaluating,komorowski2018artificial,sutton2018reinforcement}.

This work highlights the importance of discretization decisions and sequence
length in real-time causal estimation settings and suggests several future
research directions. First, this work is limited to binary treatment sequences with a single 0-to-1 transition and the results in Sections \ref{sec:synthetic_experiments} and \ref{sec:discretization_bias_ir} are limited by the Markov assumption. Additional work is necessary to investigate scenarios that do not have these limitations. Second, it would be practically useful to develop general
diagnostic techniques to guide discretization decisions when domain knowledge
is not readily available and, potentially, data driven methods for automatically
selecting the bin width to minimize estimation error. Additionally,
when continuous-time or high-resolution data is available, it may be possible
to quantify or bound the amount of discretization bias under certain
assumptions. Finally, given the observed sensitivity of these methods to bin width, it is important to continue development of continuous-time methods
that obviate the need for discretization in settings with insufficient domain knowledge to guide
bin width decisions.

\section*{Acknowledgements}
This work would not have been possible without the prior work and valuable
input of Katharine Henry. This work was supported by funding from NSF SCH
number 1418590 and a grant from the Gordon and Betty Moore Foundation number
3186.01. This information or content and conclusions are those of the authors
and should not be construed as the official position or policy of, nor should
any endorsements be inferred by NSF or the U.S. Government. Dr. Saria is a
founder of and holds equity in Bayesian Health; she is the scientific advisory
board member for PatientPing; and she has received honoraria for talks from a
number of biotechnology, research, and healthtech companies. This arrangement
has been reviewed and approved by the Johns Hopkins University in accordance
with its conflict of interest policies.

\bibliographystyle{plainnat}
\bibliography{references}

\appendix
\section{Complete descriptions of IR and TMLE}

\subsection{Iterative regression (IR)}
The IR algorithm iteratively estimates each $Q_t$, beginning with $Q_T$ and progressing backwards in time. The full algorithm is described below.

\begin{enumerate}
	\item Estimate the conditional expectation $Q_{T} =
\mathbb{E}[Y|\bar{A}_T=\bar{a}_T,\bar{L}_T,V]$ by regressing $Y$ onto $(\bar{L}_T,V)$ using data from participants who followed $\bar{a}$ through time $T$. For each participant $i$ who followed $\bar{a}$ through time $T-1$, let $\hat{Q}_{i,T}$ be this estimate evaluated at $(\bar{L}_{i,T},V_i)$.
	\item For $t = T-1,...,1$:
	\begin{enumerate}
	    \item Estimate the conditional expectation $Q_{t} = \mathbb{E}[Q_{t+1}|\bar{A}_{t}=\bar{a}_t,\bar{L}_{t},V]$ by regressing the values $\hat{Q}_{t+1}$ (calculated in the previous iteration) onto $(\bar{L}_t,V)$ using data from participants who followed $\bar{a}$ through time $t$. For each participant $i$ who followed $\bar{a}$ through time $t-1$, let $\hat{Q}_{i,t}$ be this estimate evaluated at $(\bar{L}_{i,t},V_i)$.
	\end{enumerate}
	\item Return $\hat{\psi}_{IR} = \frac{1}{N}\sum_i \hat{Q}_{i,1}$. 
\end{enumerate}

\subsection{Targeted minimum loss-based estimation (TMLE)}
The TMLE algorithm modifies the IR algorithm to add an additional \textit{targeting} step, which updates the initial estimate of $Q_t$ using the probability of treatment $g$ as detailed below.

\begin{enumerate}
	\item Estimate the probability of treatment $\hat{g}$ as in Section 4.1 of the main paper.
% 	\item Regress $Y$ onto $(\bar{L},\bar{A},V)$ using data from all participants who followed $\bar{a}$ through time $T$. For each participant who followed $\bar{a}$ through time $T-1$, estimate $\hat{Q}_{i,T}$ as the fitted regression evaluated at $(\bar{L}_i,\bar{a},V_i)$.
	\item Estimate the conditional expectation $Q_{T} =
\mathbb{E}[Y|\bar{A}_T=\bar{a}_T,\bar{L}_T,V]$ by regressing $Y$ onto $(\bar{L}_T,V)$ using data from participants who followed $\bar{a}$ through time $T$. For each participant $i$ who followed $\bar{a}$ through time $T-1$, let $\hat{Q}_{i,T}$ be this estimate evaluated at $(\bar{L}_{i,T},V_i)$.
	\item Regress $Y$ onto an intercept with $\hat{Q}_{T}$ as a fixed offset and $\frac{\mathbb{I}[\bar{A}_T = \bar{a}_T]}{\hat{g}(\bar{a};\bar{L}_T,V)}$ as
observation weights using data from participants who followed $\bar{a}$ through time $T$. For each participant $i$ who followed $\bar{a}$ through time $T-1$, let $\hat{Q}^*_{i,T}$ be this fitted regression evaluated with offset $\hat{Q}_{i,T}$.
	\item For $t = T-1,...,1$:
	\begin{enumerate}
% 		\item Regress $\hat{Q}^*_{t+1}$ onto $(\bar{L}_{t},\bar{A}_{t},V)$ using data from all participants who followed $\bar{a}$ through time $t$. For each participant who followed $\bar{a}$ through time $t-1$, estimate $\hat{Q}_{i,t}$ as the fitted regression evaluated at $(\bar{L}_{i,t},\bar{a}_{t},V_i)$.
		\item Estimate the conditional expectation $Q_{t} = \mathbb{E}[Q_{t+1}|\bar{A}_{t}=\bar{a}_t,\bar{L}_{t},V]$ by regressing $\hat{Q}^*_{t+1}$ (calculated in the previous iteration) onto $(\bar{L}_t,V)$ using data from participants who followed $\bar{a}$ through time $t$. For each participant $i$ who followed $\bar{a}$ through time $t-1$, let $\hat{Q}_{i,t}$ be this estimate evaluated at $(\bar{L}_{i,t},V_i)$.
	    \item Regress $\hat{Q}^*_{t+1}$ onto an intercept with $\hat{Q}_{t}$ as a fixed offset and $\frac{\mathbb{I}[\bar{A}_{t} = \bar{a}_{t}]}{\hat{g}(\bar{a}_{t};\bar{L}_{t},V)}$ as observation weights using data from participants who followed $\bar{a}$ through time $T$. For each participant $i$ who followed $\bar{a}$ through time $t-1$, let $\hat{Q}^*_{i,t}$ be this fitted regression evaluated with offset $\hat{Q}_{i,t}$.
	\end{enumerate}
	\item Return $\hat{\psi}_{TMLE} = \frac{1}{N}\sum_i \hat{Q}^*_{i,1}$.
\end{enumerate}

\section{Simulated RCT}
\label{app:rct}

In this section, we compare the estimators presented in Section
4 of the main paper using a modified version of the data generating
mechanism from Section 5 of the main paper where $A_t$ is sampled
without confounding (i.e., $\boldsymbol{\gamma}_1 = \mathbf{0}$ and $\boldsymbol{\gamma}_2  = \mathbf{0}$). This
simulates a randomized trial where subjects are assigned a random treatment
start time. All other details of the data generating mechanism, experimental setup, and estimation procedures remained the same. Results are shown in Figure \ref{fig:rct}. Of note, for large $\delta$, we observe
increased bias for all estimators, despite the lack of confounding.
% This is consistent with the
% analysis in Section \ref{sec:discretization_bias_ir} which does not rely on the
% presence of feedback between treatment and covariates.

\ifinlinefigs
\begin{figure}[t!]
\input{sections/fig4}
\end{figure}
\fi

\section{Variance reduction strategies}
\label{sec:variance_reduction_methods}
\label{sec:variance_reduction}

In this section, we consider the impact of three approaches --- weight clipping, pooling across treatment regimes, and pooling across time points --- to reduce estimator variance in our simulated data experiments from Section 5. We begin by introducing these approaches and describing how they change our experimental setup. Other than the changes described below, no changes were made to the data generating mechanism, assumptions, or estimation procedures described in Section 5 of the main paper.

\subsection{Weight clipping}

One simple, yet effective method for reducing the variance of estimators based on
inverse probability weighting is to clip the weights so that they fall within a
prespecified range \citep{cole2008constructing,lee2011weight}. This method prevents small sets of instances from
dominating the estimate, thereby reducing variance, but may introduce additional bias. In
our evaluation, we consider a percentile based clipping where, for a
prespecified percentile $\alpha$, the weights $\frac{1}{\hat{g}(\bar{a},\bar{L}_i,V_i)}$ among participants who followed $\bar{a}$ are clipped to be between the
$\alpha$ and $100-\alpha$ percentiles of the empirical distribution of weights among participants who followed $\bar{a}$.

\subsection{Pooling}

Another approach to reducing variance is to pool data when estimating $g$ and $Q$. We consider two types of pooling. First, we consider pooling data from multiple treatment
regimes when estimating $Q$ \citep{petersen2014targeted,tran2019double}. Specifically, to estimate $Q_t = \mathbb{E}[Q_{t+1} \mid \bar{A}_t=\bar{a}_t,\bar{L}_t,V]$ in the IR procedure, we instead estimate $\mathbb{E}[Q_{t+1} \mid \bar{A}_t,\bar{L}_t,V]$ using data from \textit{all} participants. Then, for participants who followed $\bar{a}$ through time $t-1$, we set $\hat{Q}_{i,t}$ to this estimate evaluated at $(\bar{a}_t,\bar{L}_{i,t},V_i)$. This same modification can also be made to the initial untargeted estimate of $Q_t$ in the TMLE procedure. In our simulated data experiments, this means modifying the the parametric model for $Q_t$ to be 
\begin{align}
    q_t(\bar{A}_{t},\bar{L}_{t},V;\boldsymbol{\nu}) = \nu_{t,0} + \boldsymbol{\nu}_{t,1}L_t + \boldsymbol{\nu}_{t,2}V + \boldsymbol{\nu}_{t,3}\bar{A}_t.
\end{align}
where $\boldsymbol{\nu}_t$ is estimated using data from all participants. Second, we consider pooling across time points when estimating $g$ \citep{hernan2000marginal,hernan2010causal}. This means modifying our model for $g_t = P(A_t = 1 \mid A_{t-1} = 0, \bar{L}_t, V)$ to be
\begin{align}
    m_t(\bar{L}_{t},V;\boldsymbol{\theta}) = expit(\theta_{0} + \boldsymbol{\theta}_{1} V + \boldsymbol{\theta}_{2} L_{t}).
\end{align}
where $\boldsymbol{\theta}$ is estimated using data from all participants $i$ and all time points $t$ such that $A_{i,t-1} = 0$.

% This pooling is trivial when applied to estimation of $g$, but requires more work when applied to the estimation of $Q$. Let $Q_{t,t'}$ be defined as in section \ref{sec:ir}, but with outcome variable $Y_t$. That is,

% \begin{align*}
%     Q_{t,t+1} &= \mathbb{E}[Y_t|\bar{A}_{t-1}=\bar{a}_{t-1},\bar{L}_{t-1}=\bar{l}_{t-1}]\\
%     Q_{t,t'+1} &= \mathbb{E}[Q_{t,t'}|\bar{A}_{t'-1}=\bar{a}_{t'-1},\bar{L}_{t'-1}=\bar{l}_{t'-1}]
% \end{align*}

% for $t=1,...,T$ and $t' < t$. Then, under the time-homogeneity assumption, we have $Q_{t,t'} = Q_{t-\delta,t'-\delta}$ for all $\delta <= t'$. For example, in a medical context, we might assume that, given a patient's physiological state and treatment history, future outcomes are independent of how long ago they presented at the hospital. This assumption means that we can pool across time points when estimating (untargeted) $Q$ by simultaneously estimating $\mathbb{E}[Y_t^{\bar{a}}]$ for all $t$ and pooling across $t$ at each iteration. 
\subsection{Results}
\label{app:variance_reduction_real_data}

\ifinlinefigs
\begin{figure}[t!]
\input{sections/fig5}
\end{figure}
\fi

We compared the effect of each of these variance reduction methods using the
same data generating distribution and experimental setup (aside from the modifications described above) as in Section
5 of the main paper with a discretization bin width of $\delta = \delta^* = 1$. Figure
\ref{fig:variance_reduction} shows the impact on absolute bias and variance of
each of the variance reduction approaches. Both weight clipping and temporal pooling were
extremely effective at reducing variance in the IPW and TMLE methods. In the case of IPW, weight clipping increased bias, as expected. Pooling across treatment
regimes, on the other hand, had little observable effect on either the bias or
variance of any of the estimators. 
% This may be because, in our synthetic setting, the parametric models for each $Q_t$ have 5 parameters (i.e. we have many more samples than parameters) and thus
% variance may be dominated by the effects of sequence length.

\section{Supplemental material for Section 6 of the main paper}

\ifinlinefigs
\subsection{Derivation of Equation 4 in the main paper}
\else
\subsection{Derivation of Equation 6.7 in the main paper}
\fi

As in Section 6 of the main paper, all quantities are with respect to the finest discretization, $\delta^*$, unless otherwise specified. Then, recalling that $\bar{l}_{j:k} = [l_j,l_{j+1},\dots,l_k]$ and that $L_t^{\bar{a}_{j:k}}$ is the potential outcome of $L_t$ when only the treatment subsequence $\bar{a}_{j:k}$ is set, we can rewrite the \textit{uncoarsened g-computation formula} as
\begin{align}
    \label{eq:app_po_gform_l1}
	\mathbb{E}[Y^{\bar{a}}] &= \sum_{\bar{l}} \mathbb{E}[Y \mid \bar{a}, l_{T^*}] \prod_{t=1}^{T^*} P(L_{t} = l_{t} \mid \bar{a}_{t-1}, l_{t-1})\\
	\label{eq:app_po_gform_l2}
	\begin{split}
	&= \sum_{\bar{l}(\delta)} \sum_{\bar{l}_{s(1)+1:s(2)-1}} \dots \sum_{\bar{l}_{s(T-1)+1:s(T)-1}} \mathbb{E}[Y \mid \bar{a}, l_{T^*}] P(L_1 = l_1)\\ &\mspace{270mu}\cdot \prod_{k=2}^{T(\delta)} \left\{\prod_{t = s(k-1)+1}^{s(k)} P(L_{t} = l_{t} \mid \bar{a}_{t-1}, l_{t-1})\right\}
	\end{split}\\
	\label{eq:app_po_gform_l3}
% 	&= \sum_{\bar{l}(\delta)} \mathbb{E}[Y \mid \bar{a}, l_{T}] P(L_1 = l_1) \prod_{k=2}^{K} \sum_{\bar{l}_{s(k-1)+1:s(k)-1}} P(L_{s(k)} = l_{s(k)} \mid \bar{a}_{s(k)-1}, l_{s(k) - 1}) \prod_{t = s(k-1)+1}^{s(k)-1} P(L_{t} = l_{t} \mid \bar{a}_{t-1}, l_{t-1})\\
	&= \sum_{\bar{l}(\delta)} \mathbb{E}[Y \mid \bar{a}, l_{T^*}] P(L_1 = l_1) \prod_{k=2}^{T(\delta)}\,\, \sum_{\bar{l}_{s(k-1)+1:s(k)-1}} \left\{\prod_{t = s(k-1)+1}^{s(k)} P(L_{t} = l_{t} \mid \bar{a}_{t-1}, l_{t-1})\right\}\\
	\label{eq:app_po_gform_l4}
	&= \sum_{\bar{l}(\delta)} \mathbb{E}[Y \mid \bar{a}, l_{T^*}] P(L_1 = l_1) \prod_{k=2}^{T(\delta)} P(L^{\bar{a}_{s(k-1)+1:s(k)-1}}_{s(k)} = l_{s(k)} \mid \bar{a}_{s(k-1)}, l_{s(k-1)})\\
	\label{eq:app_po_gform}
	\begin{split}
	&= \sum_{\bar{l}(\delta)} \mathbb{E}[Y \mid \bar{a}, l(\delta)_{T(\delta)}] P(L(\delta)_1 = l(\delta)_1)\\ &\qquad\qquad\qquad\qquad\cdot\prod_{k=2}^{T(\delta)} P(L(\delta)^{\bar{a}_{s(k-1)+1:s(k)-1}}_{k} = l(\delta)_{k} \mid \bar{a}_{s(k-1)}, l(\delta)_{k-1}),
	\end{split}
\end{align}
where the sum over $\bar{l}$ in Equation \ref{eq:app_po_gform_l1} indicates a sum over all possible values of the uncoarsened feature sequence; Equation \ref{eq:app_po_gform_l2} partitions the sum over $\bar{l}$ into the sum over the  components in $\bar{l}(\delta)$ (i.e., those in the coarsened sequence: $s(1),\dots,s(T(\delta))$) and  over the remaining components; Equation \ref{eq:app_po_gform_l3} follows from factorizing the nested sums in Equation \ref{eq:app_po_gform_l2} and the assumption that $s(1)=1$ and $s(T(\delta))=T^*$; Equation \ref{eq:app_po_gform_l4} follows by recognizing that (by the Markov assumption) the inner sum in Equation \ref{eq:app_po_gform_l3} is an instance of the g-computation formula applied to ``outcome" $\mathbb{I}[L_{s(k)}=l_{s(k)}]$ setting treatments at each time $s(k-1)+1$ through $s(k)-1$ and conditioning on the (uncoarsened) history through $s(k-1)$; and Equation \ref{eq:app_po_gform} follows from the observation that $L_{s(k)} = L(\delta)_k$. 

\subsection{Effect delay in the TMLE model}
\label{app:tmle_effect_delay}

Figure \ref{fig:tmle_effect_delay} shows the impact of changing the effect delay $\omega$ on the absolute bias of the TMLE estimator for different bin widths $\delta$. See Section 6 of the main paper for experimental details.

\ifinlinefigs
\begin{figure}[t!]
\input{sections/fig6}
\end{figure}
\fi

\section{Supplemental material for Section 7 of the main paper}

\subsection{List of features}
As baseline features, we
included age, sex, Charlson Comorbidity Index (CCI), hospital ID, and
indicators for comorbidities measured using billing codes including diabetes
without complications, diabetes with complications, malignant tumor, and
metastatic solid tumor. As time-varying features, we included indicators of death or discharge, temperature,
systolic blood pressure, white blood cell count, lactate, mean arterial
pressure (MAP), heart rate, respiration rate, Glasgow Coma Scale (GCS), weight,
indicators for the presence of measurements of each of the preceding
labs/vitals, and indicators for presence of stroke, seizure, dementia, fall,
gastro-intestinal (GI) bleeding, cardiac arrest, sickle cell anemia, and drug
overdose. Finally, we included as time-varying features indicators of the nine organ dysfunctions from \cite{henry2019comparison} which were summed to measure total organ dysfunction.

\subsection{Details for the parametric models used to estimate \texorpdfstring{$g$}{g} and \texorpdfstring{$Q$}{Q}}

The model for $g_t$ was modified from the model in Section 5 of the main paper to incorporate the assumption that a patient cannot initiate antibiotics or become newly censored after death or discharge. Let $D_t$ be a binary time-varying feature indicating if a patient has died or been discharged prior to time $t$. As described above, $D_t$ is included in $L_t$. Then, we estimated $g_t$ using the following parametric model
\begin{align}
    m_t(\bar{L}_{t},V;\boldsymbol{\theta}_t) =
    \begin{cases}
        expit(\theta_{t,0} + \boldsymbol{\theta}_{t,1} V + \boldsymbol{\theta}_{t,2} L_{t}) & \text{if $D_t = 0$}\\
        1 & \text{if $D_t = 1$},
    \end{cases}
\end{align}
where $\theta_t$ was estimated using data from patients who had not died, been discharged, become censored, or received antibiotics by time $t$. We used a parametric model of the same form as $m_t$ to estimate the probability of becoming newly censored at time $t$. The parametric model for $Q_t$ was modified from the model presented in Section 5 of the main paper to incorporate the assumption that a patient's outcome cannot change after death or discharge. We estimated $Q_t$ using the following parametric model
\begin{align}
    q_t(\bar{L}_{t},V;\boldsymbol{\nu}_t) = 
    \begin{cases}
    \nu_{t,0} + \boldsymbol{\nu}_{t,1}L_t + \boldsymbol{\nu}_{t,2}V & \text{if $D_t = 0$}\\
    Y & \text{if $D_t = 1$},
    \end{cases}
\end{align}
where $\boldsymbol{\nu}_t$ was estimated using data from patients who followed $\bar{a}$ through time $t$ and had not died, been discharged, or become censored by time $t$. 

\subsection{Variance reduction strategies in clinical data}

\ifinlinefigs
\begin{figure}[t!]
\input{sections/fig7}
\end{figure}
\fi

In this section, we evaluate the impact of each of each of the variance reduction approaches described in Section \ref{sec:variance_reduction_methods} of the supplementary materials on the clinical decision making problem described in Section 7 of the main paper. For each estimator and variance reduction method, we estimated the ATE and 95 \% confidence interval using a discretization bin with of $\delta = \delta^* = 1$ hour. For the weight clipping method, we used $\alpha = 0.1$. All other experimental and estimation details were exactly the same as in Section 7 of the main paper. Figure \ref{fig:hcgh_var} shows the point estimates and 95\% confidence
intervals when each of the variance reduction methods is applied. In general,
these variance reduction methods had much less of an effect in a real data
setting than they did on simulated data. Weight clipping had the largest effect
on confidence interval width for the IPW and TMLE estimators, reducing them
both from $0.50$ to $0.41$ and $0.40$ to $0.36$, respectively, without a major
shift in the point estimates. Additionally, treatment pooling, led to a small
reduction in confidence interval width, from $0.28$ to $0.26$, for the IR
estimator. Unlike in the simulated data, temporal pooling, led to an
\emph{increase} in confidence interval width for the IPW estimator, as well as
a shift in the point estimates for both IPW and TMLE (the shift in the point estimates for the
individual expected potential outcomes was even more pronounced). This is possibly due
to observable shifts in antibiotics administration behavior over time which
violates the assumptions made when pooling across time.

% \begin{figure}[!p]
%     \input{sections/fig1}
% \end{figure}

% \begin{figure}[!p]
%     \input{sections/fig2}
% \end{figure}

% \begin{figure}[!p]
%     \input{sections/fig3}
% \end{figure}

% \begin{table}[!p]
% \input{sections/table1}
% \end{table}

\end{document}